\documentclass[showpacs,prl,twocolumn,aps,superscriptaddress,preprintnumbers,letterpaper]{revtex4}
\usepackage{amsmath,amssymb}
\usepackage{epsfig}
\usepackage{graphicx}
\usepackage{amsmath}
\usepackage{amsfonts}
\usepackage{epstopdf}
\newcommand{\be}{\begin{equation}}
\newcommand{\ee}{\end{equation}}
\newcommand{\bear}{\begin{eqnarray}}
\newcommand{\eear}{\end{eqnarray}}
\newcommand{\ba}{\begin{array}}
\newcommand{\ea}{\end{array}}

\begin{document}

\title{Anomalous Chiral Fermi Surface}

%\author{Dmitri E. Kharzeev}
%\email{Dmitri.Kharzeev@stonybrook.edu}
%\affiliation{Department of Physics and Astronomy, Stony Brook University, Stony Brook, New York 11794-3800, USA}
%\affiliation{Department of Physics,
%Brookhaven National Laboratory, Upton, New York 11973-5000, USA}

%\author{Ho-Ung Yee}
%\email{hyee@tonic.physics.sunysb.edu}
%\affiliation{Department of Physics and Astronomy, Stony Brook University, Stony Brook, New York 11794-3800, USA}

\author{Ismail Zahed}
\email{zahed@tonic.physics.sunysb.edu}
\affiliation{Department of Physics and Astronomy, Stony Brook University, Stony Brook, New York 11794-3800, USA}

%%%%%%%%%
\date{\today}

\begin{abstract}
We provide a geometrical argument for the emergence of a Wess-Zumino-Witten (WZW) term in a Fermi
surface threaded by a Berry curvature. In the presence of external fields, the gauged WZW term yields a
chiral (triangle) anomaly for the fermionic current at the edge of the Fermi surface. Fermion number is conserved
though since Berry curvatures occur always in pairs with opposite (monopole) charge. The anomalous 
vector and axial currents for a a fermionic fluid at low temperature threaded by pairs of Berry curvatures are discussed.
The leading temperature correction to the chiral vortical effect in a slowly rotating Fermi surface threaded by a 
Berry curvature maybe tied to the gravitational anomaly.
\end{abstract}
\pacs{03.65.Vf,11.30.Rd,11.15Yc}

\maketitle

\setcounter{footnote}{0}

%\baselineskip 18pt \pagebreak
%\renewcommand{\thepage}{\arabic{page}}
%\tableofcontents
%\pagebreak

\vskip0.2cm

{\bf 1.\,\,Introduction.}
Chiral anomalies play a key role in QCD at low energy where for instance the chiral triangle
anomaly causes the $\pi^0\rightarrow\gamma\gamma$ decay.
Chiral anomalies have a geometrical origin that is concisely captured by Wess-Zumino-Witten (WZW) terms 
\cite{WESS,WITTEN}.  Anomalies in fermionic systems at finite density were noted in~\cite{CFL}.
Similar anomalies have emerged in dense holography~\cite{ADS,ANTON} which have a general 
explanation from entropy constrained hydrodynamics~\cite{SON1}, and more generically
WZW effective actions with matter~\cite{MISHA,SSZ}.  The role of the WZW term in chiral fluids
including the role of the gravitational anomaly, was recently analyzed in~\cite{NAIR}.
Berry phases in fermionic systems are intricatly
related to WZW-like terms. Pertinent examples can be found in the chiral bag models of hadrons
\cite{CHIRAL}.

Recently, a triangle anomaly was argued to take place in Fermi surfaces with Berry curvatures
\cite{SON}, a point of relevance to Weyl semi-metals~\cite{WEYL} and possibly graphene~\cite{GRAPHENE}. A careful but involved
analysis of the phase space structure of Fermi liquids in the presence of a Berry curvature has revealed
a triangle anomaly in the fermionic current that tags on the monopole charge of the Berry phase. Since
Berry curvatures always occur in dual pairs, charge conservation is of course garenteed.  The interplay between
a Berry phase and the chiral anomaly in QCD was originally noted in~\cite{ALEX}. The relevance of the
chiral anomaly to condensed matter physics with numerous applications to $^3$He-A was also
emphasized in~\cite{VOLOVIK}  (and references therein).

In this note, we show that the triangle anomaly in the fermioncic current follows from a WZW-like 
contribution of the Berry curvature after a homotopic extension of the quasiparticle adiabatic momentum
and a minimal $U(1)$ gauging. As expected, the normalization of the WZW-like action follows solely from 
geometry once the monopole charge is given. The ensuing currents at the edge of the Fermi surface are
plagued by the chiral (triangle) anomaly.  At low temperature, a slowly rotating Fermi surface threaded by 
Berry curvatures exhibit both chiral magnetic and vortical effects in the presence of external U(1) fields~\cite{DIMA}. 
The leading temperature correction to the chiral vortical effect maybe traced back to the 
gravitational anomaly~\cite{VOLOVIK,GRAVITY,NAIR}.

\vskip0.2cm
{\bf 2.\,\,Berry Curvature.}
The emergence of a Berry phase at the Fermi surface results from the occurence of a level 
crossing between a particle and a hole state. This crossing is dynamical in origin and maybe due
to an accidental zero in the momentum dependent gap function for weakly coupled BCS metals
or some intricate dynamics in the particle level dispersion law as in 
Weyl semi-metals or graphene for instance. The particles and holes (quasiparticles) near the zero
act as slow variables in the presence of an induced Berry phase generated by the rest of the Fermi surface acting
as fast variables.  Near the zero or crossing point, the fermion dispersion relation is linear causing the fermions to 
behave nearly relativistically. As a result, the dynamics near the Fermi surface is effectively 1+1 dimensional making
the set up ideal for the emergence of chiral anomalies and potentially the general lore of anomalous bosonization.

For simplicitly consider a level crossing at the edge of a Fermi surface 
with canonical degeneracy 2, making the induced Berry curvature
a monopole of charge $g=1/2$ (see below). If we denote by $\vec{\cal A}(\vec{p})$ the pertinent 
Berry phase or curvature near such a zero or level crossing, then it couples to a particle or a hole 
with (adiabatic) momentum ${\vec p}(t)$ through

\be
{\bf S}_B=\int\,dt\,\vec{\cal A}(\vec{p})\cdot \dot{\vec{p}}
\label{B1}
\ee
Since the Berry curvature is monopole-like and Abelian it is threaded by a Dirac string and the action 
(\ref{B1}) in 0+1 dimensions is non-local. It can be made local by extending it homotopically to 1+1 
dimensions following the original arguments in~\cite{WITTEN}. For that we define

\be
\vec{p}(t)\rightarrow k_F\,\hat{p}(t,s)
\label{B2}
\ee
with $k_F$ the Fermi momentum, $\hat{p}(t,0)=\hat{e}$   some fixed unit vector and  $\hat{p}(t,1)=\hat{p}$.  The manifold 
${\cal D}=S^1\times [0,1]$ can be regarded as the upper cup of $S^2$ with $\overline{\cal D}$ its
complement, i.e. $S^2={\cal D}+\overline{\cal D}$. While $\Pi_{0+1}(S^2)=0$, $\Pi_{1+1}(S^2)={\bf Z}$.
The latter is characterized by the topological charge or winding density

\be
{\bf W}(\hat{p})=\frac 1{8\pi}\,\epsilon^{ij}\epsilon^{abc}
\hat{p}^a\partial_i\hat{p}^b\partial_j\hat{p}^b
\label{B3}
\ee
with the labeling $i,j=0,5$,  and $a,b,c=1,2,3$, the coordination $x_0=t$ and $x_5=s$.
(\ref{B3}) carries  unit normalization

\be
\int_{S^2}\,d^2x\,{\bf W}(\hat p)=\frac 1{4\pi}\int_{S^2}\hat{p}\cdot\left(d\hat{p}\times d\hat{p}\right)=1
\label{B4}
\ee

The Berry curvature in (\ref{B1}) is a monopole flux

\be
{\bf S}_B=\int_{\cal C=\partial {\cal D}}\vec{\cal A}(\vec{p})\cdot d{\vec{p}}\equiv \int_{\cal{D}}\,d{\cal A}(\vec{p})
\label{B6}
\ee
with the form notation subsumed. A Berry curvature of unit flux through the Fermi surface amounts to

\be
\int_{S^2={\cal D}+\overline{\cal D}}\,d{\cal A}=g\,\int_{S^2}\,\hat{p}\cdot(d \hat{p}\times d{\hat{p}})=2\pi
\label{B7}
\ee
which is effectively a hedgehog monopole in momentum space of charge $g=1/2$ centered in the Fermi sea.
Therefore, the non-local form (\ref{B1}) in 0+1 dimensions can be rewritten in local form in 1+1 dimensions through

\be
{\bf S}_B=4\pi g\,\int_{\cal D}\,d^2x\,{\bf W}(\hat p)=\frac{g}{2k_F^3}\int_{\cal D}d^2x\,
\epsilon^{ij}\epsilon^{abc} {p}^a\partial_i{p}^b\partial_j{p}^b
\label{B8}
\ee
after relaxing the form notation. (\ref{B8}) is the WZW term for the Berry curvature of flux 1 or
monopole charge $g=1/2$ centered in a Fermi sphere of radius $k_F$. (\ref{B8}) is also well
established in condensed matter~\cite{VOLOVIK,HASAN,QI}.

A Berry curvature of flux $k$ through the Fermi surface amounts to a monopole of charge $g=k/2$ centered in the Fermi sphere,
while a Berry curvature of flux $-k$ amounts to an antimonopole of charge $g=-k/2$. A monopole will cause the particle and
hole excitations at the zero or level crossing to be right handed say, while an antimonopole will cause them to left handed.
The labelling right and left is conventional. In materials level crossings occur always in pairs making the net flux always zero.

\vskip0.2cm
{\bf 3.\,\,Chiral Anomaly.}
To assess the effect of the Berry curvature (\ref{B8}) on the transport of the fermion number current
around the zero or level crossing in the presence of external electromagnetism and at the edge of
the Fermi surface, we need to gauge (\ref{B8}).  For that we note that if we were to formally extend

\be
p^a(t,s)\rightarrow p^\mu(t,s,\vec{x}) 
\label{B9}
\ee
for the sake of the argument (we will revert to $p^a(t,s)$ shortly) 
, which is seen to be valued in ${\cal D}\times R^3$,
 then (\ref{B8}) turns to a Chern-Simons-like  contribution in 5 dimensions

\be
{\bf S}_{\perp B}\equiv N_\perp\,{\bf S}_B=\frac{g}{2k_F^3}\,\frac{N_\perp}{V_3}
\int_{{\cal D}\times R^3}\, p\,(dp)^2
\label{B10}
\ee
for $N_\perp$ quasiparticles. 
Here $N_{\perp}/V_3=n_\perp k_F$ and $n_\perp=k_F^2/2\pi^2$ denotes the fermion density at the Fermi surface for
a single fermion species.  Near a zero or level crossing, the quasiparticles 
contribute {\it coherently} to the Chern-Simons-like term, thus the multiplication by $N_\perp$. Their  number density is given by the density $n_\perp$ at the Fermi surface.

It is now straightforward to gauge (\ref{B10}), say by minimal substitution $p\rightarrow p+A$. Thus, the gauged
WZW term in form notation is

\be
{\bf S}_{\perp B}=\frac{gn_\perp}{2k_F^2}\,
\int_{{\cal D}\times R^3}\, (p+A)\,\left(dp+\frac 12 F\right)^2
\label{B11}
\ee
We now revert to $p^\mu\rightarrow p^a(t,s)$ for our Berry curvature or hedgehog monopole. 
$F$ is the U(1) field strength. From (\ref{B11}) 
it follows that the Fermionic current is anomalous. Indeed, in the presence of a U(1) gauge field $A$ the 
fermionic current carried by the particle and hole excitations 
at the level crossing or zero can be thought of as either right-handed (Berry curvature with
net positive flux) or left handed (Berry curvature with net negative flux).  Both currents couple {\it normally} to the 
U(1) gauge field through

\be
{\bf S}_{R,L}=\int_{R^4}\,{\bf J}_{R,L}A
\label{B12}
\ee
The Noether construction $A\rightarrow A+d\xi$ shows that

\be
d\,{\bf J}_R=\frac{gn_\perp }{2k_F^2}\left(dp+\frac 12 F\right)^2
\label{B13}
\ee
is anomalous.
A similar relation holds for ${\bf J}_L$ with $g\rightarrow -g$.
In (\ref{B13}) $p^\mu\equiv p^a(t,s=1)$ (after reverting to $p^a(t,s)$)
and its contribution has no support by the antisymmetric contraction.
Thus

\be
d\,{\bf J}_R=\frac{gn_\perp }{8k_F^2}\, F^2\equiv \frac{g}{2\pi^2}\,\vec{E}\cdot\vec{B}
\label{B14}
\ee
after using the explicit contribution of the single species fermions at the Fermi surface. For the 
Berry curvature of general charge $g=k/2$ or $k$-fluxes through the Fermi surface, this is the result
established recently in~\cite{SON} using Fermi liquid theory and transport arguments and earlier
 in~\cite{VOLOVIK} in the context of $^3$He-A.

Since at the zero
or level crossing the particle-hole excitations are effectively Weyl, (\ref{B14}) is the expected
anomaly for a free Weyl fermion in 1+3 dimensions. Note that the features of the Fermi surface dropped
out of (\ref{B14}) as expected from geometry. Anomalies are infrared manifestations of ultraviolet physics
that are insensitive to matter~\cite{BARDEEN}.

In many ways, (\ref{B14}) carries the essentials of the anomaly matching condition in matter whereby 
the infrared degrees of freedom at the Fermi surface are mapped onto the ultraviolet character of
the chiral anomaly in the vaccum. 

It is worth noting that when the number of quasiparticles in (\ref{B10}) encompasses all of the Fermi surface
then $N_\perp/V_3=(n_\perp k_F)/3$ is just the Fermi density, in which case 

\be
d\,{\bf J}_R=\frac{gn_\perp }{24k_F^2}\, F^2\equiv \frac{g}{6\pi^2}\,\vec{E}\cdot\vec{B}
\label{B14X}
\ee
(\ref{B14}) and (\ref{B14X}) are the covariant  and consistent  form of the Abelian chiral anomaly respectively,
both of which are known to follow from specific UV regularizations for fermions in the vacuum~\cite{DINE}.  
Here, they are realized through a different counting of states either at the surface (\ref{B14}) or through the bulk (\ref{B14X})
of the Fermi surface. Both may have a realization in materials differing by the degree of coupling of the quasiparticles.

\vskip0.2cm
{\bf 4.\,\,Anomalous fermionic fluid.}
If we were to treat the quasi-particle excitations at the Fermi surface as a fluid with a 
(momentum dependent) fluid velocity $v^\mu$ and
a chemical potential $\mu_R$ (following the right convention for a zero or crossing with positive flux), then
chiral magnetic and vortical contributions are expected. 

In the presence of anomalies the chemical potentials
are defined using the conserved but gauge fixed right and left currents. We note that the gauge fixing in the vector
current is redundant since it is conserved (see below). In leading
order they follow readily  by the substitution $A\rightarrow A+\mu_{R,L}\,v$ in the WZW term
or the anomaly contribution~\cite{MISHA}. Specifically, (\ref{B14}) now reads (for the right current)

\begin{eqnarray}
d\,{\bf j}_R=&&\frac{g }{16\pi^2}\, (F+2\mu_R\,dv)^2\nonumber\\
=&&\frac {g}{16\pi^2} \left(F^2+4\mu_R\,F\,dv+4\mu_R^2(dv)^2\right)
\label{B15}
\end{eqnarray}
and ${\bf j}_R={\bf n}_R v$ now the {\it normal} constitutive current with quasiparticle density ${\bf n_R}$. (\ref{B15}) 
can be reshuffled in the form

\be
d\tilde{\bf j}_R\equiv d\left({\bf j}_R-\frac{g}{4\pi^2}\left(\mu_R\,F\,v+\mu_R^2\,vdv\right)\right)
=\frac{g }{16\pi^2}\, F^2
\label{B16}
\ee
(\ref{B16}) shows that the constitutive but normal current ${\bf j}_R$ acquires an anomalous contribution, the sum of which yields

\be
\tilde{\bf j}_R\equiv {\bf j}_R-\frac{g}{4\pi^2}\left(\mu_R\,F\,v+\mu_R^2\,vdv\right)
\label{B17}
\ee
which obeys the triangle or chiral anomaly on the Fermi surface.  
Similar relations hold for the left current with the substitution $g\rightarrow -g$.
The vector current $\tilde{\bf j}={\tilde{\bf j}}_R+{\tilde{\bf j}}_L$ is anomaly free, while the axial current
$\tilde{\bf j}_A={\tilde{\bf j}}_R-{\tilde{\bf j}}_L$ is anomalous. 

The first anomalous contribution in (\ref{B17}) is the chiral magnetic contribution while the second anomalous
contribution is the chiral vortical effect.  For instance the spatial vector and axial currents flowing through 
a rotating but cold Fermi surface threaded by a dual pair of Berry curvatures read

\begin{eqnarray}
\int\,\frac{d\hat{p}}{4\pi}\,\,{\tilde{\bf j}}^i&=& -\frac{g}{2\pi^2}(\mu_R-\mu_L)\,B^i+\frac{g}{4\pi^2}
(\mu_R^2-\mu_L^2)\,\omega^i\nonumber\\
\int\,\frac{d\hat{p}}{4\pi}\,\,{\tilde{\bf j}}_A^i&=& -\frac{g}{2\pi^2}(\mu_R+\mu_L)\,B^i+\frac{g}{4\pi^2}
(\mu_R^2+\mu_L^2)\,\omega^i\nonumber\\
\label{B18}
\end{eqnarray}
with $\vec\omega$ the external circular velocity. The emergence of a current in the presence of a slowly
rotating Fermi surface was also noted in~\cite{VILENKIN,ALEKSEEV,DIMAZ}.

At low temperature the axial vortical effect is expected to be shifted 

\be
\mu_{R,L}^2\rightarrow \mu_{R,L}^2+\frac{(\pi T)^2}{24}
\label{B19}
\ee
while the vector vortical effect is not. The temperature shift
appears naturally in the context of a rotating superfluid $^3$He-A system at low temperature and reflects
on the generic character of the mixing between axial and gravitational anomalies in gapless
constitutive systems~\cite{VOLOVIK,GRAVITY,NAIR}. A more microscopic description of the
constitutive currents shows that the temperature shift follows from  the leading tadpole corrections as discussed in~\cite{MISHA} through thermal phonons at low temperature in a superfluid state, whereby $\Pi^2/F^2\approx (\pi T)^2/\mu^2$ with $F^2\approx n_\perp$ at the Fermi surface.

\vskip0.2cm
{\bf 5.\,\,Conclusions.}
We have shown that the triangle anomaly established recently in~\cite{SON} follows from the
pertinent gauged WZW term associated to the Berry curvature in a Fermi surface. This result
was expected, as all anomalies are of geometrical nature and insensitive to the details of the
underlying dynamics here taking place at the Fermi surface. The origin of the Berry curvature
in a Fermi surface while intricate dynamically, is manifested by particle-hole crossing at 
specific points of the Fermi surface. Near these points, the approximate quasiparticle spectrum is effectively 
2 dimensional and relativistic.

In the presence of U(1) gauge fields, the right and left fermionic quasiparticle currents
are anomalous with the chiral or triangle anomaly being the lore.  Our analysis through
(\ref{B14}) shows how the infrared degrees of freedom at the Fermi surface are mapped onto
the ultraviolet content of the chiral anomaly. This is an example of how the anomaly matching
condition operates around a Fermi surface.  This point is of interest to dense QCD
at weak coupling whereby Fermi surfaces are expected.

A rotating Fermi fluid threaded by Berry curvatures at low temperature
exhibits both chiral and vortical effects. The leading temperature effects appear to be related to
the gravitational anomaly noted in~\cite{VOLOVIK,GRAVITY,NAIR} and perhaps generic. These leading temperature effects are
tadpole like and unambiguous in the anomalous superfluid or effective Lagrangian analysis 
in~\cite{MISHA}.  They maybe measurable through the axial vortical effect. 

Finally, it would be interesting to explore the possible occurence of {\it non-Abelian} Berry curvatures
in Fermi surfaces whereby non-Abelian anomalies can be realized.  We recall that non-abelian Berry phases 
emerge naturally in chiral bag models where 3 quarks  are (somehow) trapped, and contribute essentially to their quantum numbers
~\cite{CHIRAL}.  Non-Abelianity requires additional internal degeneracies on the particle and hole quasiparticles
at the edge of the Fermi surface with spin being an obvious candidate.
\\
\vskip0.2cm
{\bf Acknowledgements.}
I would like to thank Dima Kharzeev for bringing Ref.~\cite{SON} to my attention, 
and Gokce Basar, Gerald Dunne, Dima Kharzeev and Ho-Ung Yee for discussions. 
I also thank Grisha Volovik for several e-discussions following the appearance of
this note. This work was supported by the U.S. Department of Energy under Contract No.
DE-FG-88ER40388.

%%%%%%%%%%%%%%%%%%%%%%%%%%%%%%%%%%%%%%%%%%%%%%%%%%%%%%%%%%%%%%%%%%%

\end{document}